\documentclass[aps,pra,twocolumn,showpacs,nofootinbib]{revtex4}
\usepackage{bm,bbm,amsmath,amssymb,amsbsy,graphicx,hyperref,times,txfonts}

\usepackage[latin1]{inputenc}

\newcommand{\eq}[1]{Eq.~(\ref{#1})}
\newcommand{\ineq}[1]{Ineq.~(\ref{#1})}
\newcommand{\ket}[1]{|#1\rangle}
\newcommand{\bra}[1]{\langle#1|}

\begin{document}

\title{Quantum discord for general two--qubit states: Analytical progress}

\author{Davide Girolami}
\thanks{pmxdg1@nottingham.ac.uk}
\author{Gerardo Adesso}
\thanks{gerardo.adesso@nottingham.ac.uk}
\affiliation{School of Mathematical Sciences, University of Nottingham, University Park, Nottingham NG7 2RD, United Kingdom}

\date{April 01, 2011}

\begin{abstract}
 We present a  reliable algorithm to evaluate quantum discord for general two--qubit states, amending and extending an approach recently put forward for the subclass of $X$--states. A closed expression for the discord of arbitrary states of two qubits cannot be obtained, as the optimization problem for the conditional entropy requires the solution to a pair of transcendental equations in the state parameters. We apply our algorithm to run a numerical comparison between quantum discord and an alternative, computable measure of non-classical correlations, namely the geometric discord. We identify the extremally non-classically correlated two--qubit states according to the (normalized) geometric discord, at fixed value of the conventional quantum discord. The latter cannot exceed the square root of the former for systems of two qubits.
\end{abstract}


\maketitle

 \section{Introduction}
 It is as well theoretically proven as experimentally tested that the exploitation of entangled states improves our ability to manage information in several ways \cite{hororev}.  The development of quantum information theory is definitely due to this fundamental result \cite{nielsen}.  However, it has been recently discovered that even multipartite separable states can play a relevant role in performing better-than-classical communication and information protocols  \cite{dattaqc,piani,singapore}.  In general, then, the usefulness of a particular state of a quantum system for such tasks can be traced back to the presence of internal correlations  among parts of the system, i.e., to the knowledge that an observer Alice, probing a subsystem $A$, gains about the state of another subsystem $B$, controlled by another observer Bob, and vice versa.

  The quantitative and qualitative evaluation of such correlations and the proper discrimination of their nature --- classical versus quantum --- stand as open problems. Several quantifiers of non-classicality of  correlations have been introduced in literature \cite{zurek,HV,terhal,mid,deficit,moelmer,modi,dakic,gap,rossignoli} (not to be confused with the non-classicality of quantum optical states \cite{nonclasscorr}), but  there are still neither clear criteria of faithfulness or {\it bona-fide}-ness for them, nor a well established hierarchy of reliability. In spite of that, they are heavily used in current research \cite{usediscord,usemid,ferraro,discordGauss,operdiscord,mista,provapovm,luo,alber,cinesi_sensati,nuovodiscord,james,galve,luofu}. The most popular one is by far the \emph{quantum discord} \cite{zurek,HV}, a measure of non--classical correlations which goes beyond entanglement and whose definition has an immediate interpretation in information theory:  discord equals the difference of the total correlations between two subsystems $A$ and $B$ before and after a local measurement process is performed on one of them. The quantum discord admits at least three other operational interpretations, in contexts ranging from thermodynamics to communication protocols, such as the quantum state merging \cite{operdiscord}. The evaluation of the quantum discord is a hard task from a computational point of view, implying an optimization of the conditional entropy between Alice and Bob, ${\cal S}(A|B)$, over all local (generalized) measurements on one party, which is often obtainable  by numerical methods only. A closed analytical solution is known in the case of arbitrary two--mode Gaussian states \cite{discordGauss}, under the restriction of Gaussian local measurements. Narrowing our overview to two--qubit states, an analytical expression of discord has been derived in particular for the  subclass of so-called $X$--states \cite{luo,alber,cinesi_sensati,nuovodiscord}, and a successful attempt to generalize this procedure has not been advanced yet (to the best of our knowledge), apart from an upper bound quantity for the discord defined in \cite{cinesi_sensati}.

  The difficulty in calculating quantum discord motivated the introduction of alternative measures of non-classical correlations.
  In particular, the \emph{geometric discord} \cite{dakic} is one such a measure, which quantifies the amount of non-classical correlations of a state in terms of its minimal distance from the set of genuinely classical states. The geometric discord involves a simpler optimization and is easily computable analytically  for general two--qubit states.  However, its relationship with the original quantum discord is not entirely clear at the present stage.

   In this work, we present an algorithm to calculate  quantum discord for  general two--qubit states. First, we obtain an  explicit and simplified expression for the conditional entropy, exploiting the Bloch representation of the density matrix; then, we employ new variables that allow us to set the optimization conditions in a closed form. Finally, we associate them to constraints over the eigenvalues of the statistical ensemble obtained after the measurement process.  Our approach qualifies as the most efficient and reliable way to evaluate quantum discord for arbitrary states of two qubits \cite{noterespect}.

 Exploiting our algorithm, we perform a detailed numerical exploration of the Hilbert space of two--qubit states to compare quantum discord and the geometric discord as quantifiers of non-classical correlations. We shed light on the relationship between these two quantities by identifying the states that extremize geometric discord at fixed quantum discord (and vice versa). We are motivated by the aim of  establishing a reliable hierarchy of non-classical states based on physically and mathematically consistent criteria. We find that, interestingly, the quantum discord of a two--qubit state can never exceed  its (normalized) geometric discord. In analogy with  the study of maximal entanglement \cite{nemoto}, we find that the notion of maximal non-classical correlations is measure--dependent: Therefore, the feasibility in a specific experimental realization will determine which, among the various classes of maximally quantumly correlated states, are the most suitable ones to be employed for applications.

The paper  is organized as follows. Section \ref{sec1} provides an introduction to quantum discord  and geometric discord                          (hereafter denoted by ${\cal D}$ and $D_G$, respectively), presenting  definitions, main properties and  a summary of the results obtained in the literature. In Section \ref{sec2}, we pursue an analytical approach to the calculation of quantum discord for  general two--qubit states, eventually recasting the optimization problem into a system of two  transcendental equations, whose solution specifies the local measurement that minimizes the conditional entropy. Section \ref{sec3} concerns the comparison between quantum discord and geometric discord by using the results of the previous sections; we identify the classes of states with maximal and minimal geometric discord at fixed quantum discord. Finally, Section \ref{sec4} recalls the main results of our work and suggests further issues worthy of investigation.

 \section{Basic definitions}\label{sec1}
 \subsection{Quantum discord}\label{secQD}
 One of the lessons we can apprehend from Quantum Mechanics is that the measurement process disturbs the state in which a physical system is set. This differs from what happens in the classical scenario, hence it is possible to conclude that the disturbance induced by a measurement on a state is a good evidence of its ``quantumness''. Now, let us suppose to have a bipartite system in a certain state and make a measurement on one of its subsystems. We can  analyze the nature of the internal correlations of the system in such a state by studying  in which way they are affected by the measurement. Information Theory provides the  tools to accomplish this task.

  As a starting point, let us consider the case of information stored in two classical probability distributions. The quantity expressing the total amount of correlations between two random variables $X$ and $Y$, assuming values $\{x_i\},\{y_j\}$ with probability $\{p_i\},\{q_j\}$, is the \emph{mutual information} ${\cal I}$, defined as
 \begin{eqnarray}\label{ixy}
{\cal I}(X:Y)={\cal H}(X)+{\cal H}(Y)-{\cal H}(X,Y),
\end{eqnarray}
where ${\cal H}(X)=-\sum _i p_i \log_2 p_i$ is the Shannon entropy associated to the random variable $X$; consequently ${\cal H}(Y)=-\sum _j q_j \log_2 q_j$ is the entropy for $Y$ while ${\cal H}(X,Y)$ is the joint entropy of $X$ and $Y$.
Following Bayesian rules, we can retrieve equivalent formulations, ${\cal I}(X:Y)={\cal I}(Y:X)={\cal J}(X:Y)= {\cal J}(Y:X)$, where
\begin{eqnarray}
{\cal J}(X:Y)&=&{\cal H}(X)-{\cal H}(X|Y)\,,\\
{\cal J}(Y:X)&=&{\cal H}(Y)-{\cal H}(Y|X)\,.
\end{eqnarray}
 Here the conditional entropy is straightforwardly defined as ${\cal H}(X|Y)={\cal H}(X,Y)-{\cal H}(Y)$, and represents how much uncertainty (ignorance) we have on $X$ given the value of $Y$ (and  vice versa for ${\cal H}(Y|X)$).

In the quantum scenario,  we consider a bipartite system $AB$ described by a density matrix $\rho \equiv \rho_{AB}$, and subsystems $A,B$  with marginal density matrices $\rho_A, \rho_B$. The mutual information can be used once more to quantify the total  correlations between $A$ and $B$. The quantum analogue of the expression (\ref{ixy}) straightforwardly reads
\begin{eqnarray}
{\cal I}(A:B)={\cal S}(A)+{\cal S}(B)-{\cal S}(A,B),
\end{eqnarray}
 where ${\cal S}(A)=-\text{Tr}[\rho_A \log_2 \rho_A]$ is the von Neumann entropy of subsystem $A$, and equivalently ${\cal S}(B)=-\text{Tr}[\rho_B \log_2 \rho_B]$, ${\cal S}(A,B)=-\text{Tr}[\rho \log_2 \rho]$. On the other hand, if we try to define a quantum version of ${\cal J}$ we have
 \begin{eqnarray}
{\cal J}(A:B)={\cal S}(A)-{\cal S}(A|B),
\end{eqnarray}
  which is an  ``ambiguous'' quantity \cite{zurek,HV}, since the conditional quantum entropy ${\cal S}(A|B)$ is  clearly depending on which observable we have measured on $B$. Using the same notation of \cite{alber},  we recall that a von Neumann measurement (from now on just measurement) on $B$  projects the system into a statistical ensemble $\{p_k,\rho_k\}$, such that
\begin{eqnarray}\label{measure}
 \rho\rightarrow \rho_k  =\frac{(I_A\otimes P_{Bk})\rho (I_A\otimes P_{Bk} )}{p_k},
 \end{eqnarray}
 where
 \begin{eqnarray}
p_k&=& \text{Tr}[\rho (I_A\otimes P_{Bk} )] \\\nonumber
 P_{Bk}&=&V \Pi _k V^\dagger\\ \nonumber
\Pi _k &= &|k \rangle\langle k|,\ \  k=0,1\\\nonumber
V &\in& SU(2). \nonumber
\end{eqnarray}
We remind that in this case study the use of generalized positive-operator-valued-measurements (POVMs) is not required, since it has been proven in \cite{provapovm} that for two--qubit states the optimal measurement for the conditional entropy is always a projective one. We can say that the amount of truly classical correlations is expressed by  the mutual information obtained adopting the least disturbing measurement \cite{HV}:
 \begin{eqnarray}
{\cal C}(A:B)&=&\text{max}_{\{P_{Bk}\}}{\cal J}(A:B) \\
&=&{\cal S}(A)-\text{min}_{\{P_{Bk}\}}\sum_k p_k {\cal S}(A|B_{\{P_{Bk}\}}).\nonumber
\end{eqnarray}
      Consequently, the amount of genuinely quantum correlations, called \emph{quantum discord}, is given by \cite{zurek}
      \begin{eqnarray}\label{discord}
      &&{\cal D}(A:B)={\cal I}(A:B)- {\cal C}(A:B)\\\nonumber
      &&\quad = {\cal S}(B)- {\cal S}(A,B) + \text{min}_{\{P_{Bk}\}}\sum_k p_k {\cal S}(A|B_{\{P_{Bk}\}}).
      \end{eqnarray}
  We note that quantum discord is not symmetric:
  \begin{eqnarray}
  {\cal S}(A)-{\cal S}(A|B)&\neq& {\cal S}(B)-{\cal S}(B|A) ;
  \end{eqnarray}
   performing the measurement on Alice's subsystem rather than on Bob's one is perfectly legitimate, but it returns in general a different value of discord. See e.g. \cite{gap,mista} for a discussion about the implications of it. It is immediate to verify that not only entangled states, but almost all separable states have a non--vanishing quantum discord \cite{ferraro}, i.e. are affected by the measurement process, thus exhibiting some pretty quantum properties.   In the case of pure bipartite states, the discord reduces to the marginal entropy of one of the two subsystems and therefore to the canonical measure of entanglement. Quantum discord for two--qubit states is normalized to one.

  \subsection{Geometric discord}
Recently, it has been argued that the experienced difficulty of calculating quantum discord can be coped, for a general two--qubit state, with the introduction of its geometrized version, hereby just called \emph{geometric discord} \cite{dakic}.\\
Let us suppose, to be coherent with section \ref{secQD}, to have a bipartite system $AB$ and to make a measurement on $B$. As we have remarked,  almost all (entangled or separable) states are disturbed by the measurement; however, there is a subclass of states which is invariant and presents zero discord.  It is the class of the so--called \emph{classical--quantum} states \cite{piani}, whose elements have a density matrix of this form
\begin{eqnarray}\label{cq}
\rho = \sum _i  p_i \rho _{Ai}\otimes |i\rangle  \langle i | ,
\end{eqnarray}
where $p_i$ is a probability distribution, $\rho_{Ai}$ is the marginal density matrix of $A$ and  $\{|i\rangle\}$ is an orthonormal vector set. A classical--quantum state is not affected by a measurement on $B$ in any case.

 Letting $\Omega$ be the set of classical--quantum two--qubit states,  and $\chi$ be a generic element of this  set, the geometric discord $D_G$ is defined as  the distance between  the state $\rho$ and the closest classical--quantum state. In the original definition \cite{dakic}, the (squared) Hilbert--Schmidt distance is adopted. Recalling that  $||A ||_2^2=\text{Tr}[A A^T]$ is the square of the Hilbert--Schmidt norm of a matrix $A$,  the geometric discord has been introduced as
 \begin{eqnarray}\label{dgdef}
  D_G(\rho)=\text{min}_{\chi \in \Omega}||\rho -\chi ||_2^2.
  \end{eqnarray}

 It is possible to obtain an explicit closed expression of $D_G$ for two--qubit states. First, one needs to express the $4\times 4$ density matrix of a two--qubit state in the so--called Bloch basis \cite{verstraete}:
 \begin{eqnarray}\label{bloch}
  \rho&=& \frac14 \sum_{i,j=0}^3 R_{ij} \sigma_i \otimes \sigma_j \\
  \nonumber &=& \frac 14\bigg(I_{4\times 4}+\sum_{i=1}^3 x_i\sigma_i \otimes I_{2\times 2} \\ \nonumber
  & & \quad +\sum_{j=1}^3 y_j I_{2\times 2}\otimes \sigma_j+\sum_{i,j=1}^3 t_{ij} \sigma _i\otimes\sigma_j\bigg),
  \end{eqnarray}
where $R_{ij}=\text{Tr}[\rho(\sigma_i\otimes \sigma_j)]$, $\sigma_0=I_{2\times 2}$, $\sigma _i$ ($i=1,2,3$) are the Pauli matrices, $\vec{x}=\{x_i\},\vec{y}=\{y_i\}$ are the three--dimensional Bloch vectors associated to the subsystems $A,B$, and  $t_{ij}$ denote the elements of the correlation matrix $T$.  Then, it is shown in \cite{dakic} that the geometric discord is given by
 \begin{eqnarray}\label{dgformula}
 D_G(\rho)= \frac 14(||\vec y \vec y^T||_2 + ||T||_2^2 -k),
  \end{eqnarray}
  with $k$ being the largest eigenvalue of the matrix $\vec y \vec  y^T+  T^TT$ (in case of measurement on Alice, one needs to replace $\vec{y}$ with $\vec{x}$ and $T^TT$ with $TT^T$). An alternative formulation for the geometric discord has been provided in \cite{luofu}. It is easy to see that $D_G$ is not normalized to one: its maximum value is $1/2$ for two--qubit states, so it is natural to consider $2D_G$ as a proper measure for a comparison with the quantum discord ${\cal D}$.

 \section{Quantum discord for two--qubit states}\label{sec2}
 \subsection{General setting}
 Even though quantum discord has an apparently simple definition \cite{zurek}, the practice reveals that its explicit evaluation is hard to accomplish. In this paper  we restrict our attention to two--qubit states. An analytical algorithm  has been proposed for the subclass of states with maximally mixed marginals (described by five real parameters)  in \cite{luo}. Also, an extension to states spanned by seven real parameters, called $X$--states because of the peculiar form of their density matrix (with vanishing elements outside the leading diagonal and the antidiagonal), has been introduced in \cite{alber}, and amended by \cite{nuovodiscord}. Here, we attempt to generalize the procedure to the entire class of two--qubit states.

   First, we consider that performing local unitary transformations we can recast the density matrix for an arbitrary two--qubit state, Eq.~(\ref{bloch}),  in the Bloch normal form \cite{verstraete,luo}
 \begin{eqnarray}\label{simpler}
\rho&=&\frac 14(I_{4\times 4}+\sum_i a_i\sigma_i \otimes I_{2\times 2}+\sum_i b_i I_{2\times 2}\otimes \sigma_i\nonumber \\
 &+&\sum_i c_i \sigma _i\otimes\sigma_i),
 \end{eqnarray}
  that is a density matrix completely defined by nine real parameters arranged in three $3$-dimensional column vectors $\vec{a}=\{a_i\}$, $\vec{b}=\{b_i\}$ and $\vec c =\{c_i\}$.
   This follows from the fact that local unitary operations $\rho'=(U_A \otimes U_B) \rho (U_A \otimes U_B)^\dagger$ correspond to left and right multiplication of the Bloch matrix $R$ with orthogonal matrices \cite{verstraete},
   \begin{equation}\label{ortho}
   R'=\left(
        \begin{array}{cc}
          1 & 0 \\
          0 & O_A^T \\
        \end{array}
      \right) R \left(
        \begin{array}{cc}
          1 & 0 \\
          0 & O_B \\
        \end{array}
      \right)\,,
      \end{equation}
   with $O_{A,B} \in SO(3)$. It is then straightforward to obtain the normal form of Eq.~(\ref{simpler}): one needs to calculate the singular value decomposition of the lower diagonal $3\otimes3$ block $T$ of $R$, $T=O_A C O_B^T$, divide $O_A$ and $O_B$ by their respective determinant (to make sure they both have determinant $+1$), and then apply Eq.~(\ref{ortho}). The density matrix is correspondingly transformed into the normal form of Eq.~(\ref{simpler}), where the $c_i$ are identified with the elements of the diagonal matrix $C$, i.e., the singular values of $T$. Every two--qubit state can be then transformed in its simplified normal form by means of local unitaries (which preserve entanglement and correlations in general, by definition), so we can restrict our analysis to density matrices of this type  without incurring in any loss of generality.

  Now, we move to calculate the quantum discord ${\cal D}$, \eq{discord}, for generic states in normal form. The marginal entropy ${\cal S}(B)$ and the global entropy ${\cal S}(A,B)$ are trivial to obtain.  The main issue regards the optimization involved in the conditional entropy.

 \subsection{Conditional Entropy}
   Firstly, we have to write the conditional entropy in an explicit form, adopting for simplicity the notation in \cite{alber}. We remarked that a measurement sends the state $\rho$ into an ensemble $\{p_k,\rho _k\}$ as expressed in \eq{measure}.  The entropy of the ensemble can be written as
   \begin{eqnarray}
   \tilde {\cal S}&=&\sum_k p_k {\cal S}(A|B_{\{P_{Bk}\}})=\sum_k p_k {\cal S}(\rho_k)\\\nonumber
   &=&p_0 S_0 +p_1 S_1,
   \end{eqnarray}
   where $S_0, S_1$ are the entropies associated  to $\rho_0,\rho_1$.\\
   The measurement is defined by the quantity $P_{Bk}$ and is consequently parametrized by the elements of the unitary matrix $V$, which we can write in the basis of the Pauli matrices as
   \begin{eqnarray}
   V&=&v_0 I_{2 \times 2} + \vec v  \vec\sigma \\\nonumber
   &=& \left(
\begin{array}{cc}
 v_0+v_3 & v_1-iv_2  \\
 v_1+iv_2 & v_0- v_3
\end{array}
\right).
   \end{eqnarray}
    We notice that  the real vector $\{v_0,\vec v\} =\{v_i\}$ has norm one. Therefore, it is possible to rearrange the four parameters in three variables only, for example in this way,
    \begin{eqnarray}
   h&=& v_0v_1+v_2v_3\nonumber\\
j&=&v_1v_3-v_0v_2\\
    k&=&v_0^2+v_3^2.\nonumber
     \end{eqnarray}
    Setting the vectors $\vec{X} =\{2 j,2 h, 2k-1\}$ and  $\vec{m}_{\pm}=\{m_{i\pm}\}=\{a_i \pm c_i X_i\}$,  we have that,  after a straightforward calculation, the conditional entropy   takes the following  expression:
\begin{eqnarray}\label{scond}
 \tilde {\cal S}\!&\!=\!&\!-\frac 14\left\{ (1-\vec{b}\cdot\vec{X})\Bigg[\left(1-\frac{| \vec{m}_-|}{1-\vec{b}\cdot\vec{X}}\right)\log_2\!\left(1-\frac{|\vec{m}_-|}{1-\vec{b}\cdot\vec{X}}\right)\right.\nonumber\\
&+&\left(1+\frac{|\vec{m}_-|}{1-\vec{b}\cdot\vec{X}}\right)\log_2\!\left(1+\frac{| \vec{m}_-|}{1-\vec{b}\cdot\vec{X}}\right)\Bigg]\nonumber\\
 &+&(1+\vec{b}\cdot\vec{X})\Bigg[\left(1-\frac{|\vec{m}_+|}{1+\vec{b}\cdot\vec{X}}\right)\log_2\!\left(1-\frac{| \vec{m}_+|}{1+\vec{b}\cdot\vec{X}}\right)\nonumber \\
&+&\left.\left(1+\frac{| \vec{m}_+|}{1+\vec{b}\cdot\vec{X}}\right)\log_2\!\left(1+\frac{| \vec{m}_+|}{1+\vec{b}\cdot\vec{X}}\right)\Bigg]\right\}.
\end{eqnarray}

This result is consistent with the  formula provided in the Appendix of \cite{cinesi_sensati}, but we have reached here a simpler expression  by exploiting the normal form of the density matrix, \eq{simpler}.
However, we have to remark that in this picture there is still an amount of redundancy \cite{cinesi_sensati}.   A projective measurement on a two--qubit state can be characterized by  two independent variables only, identifiable as the angles $\theta$ and $\phi$, which parametrize a generic single--qubit pure state as $|\psi\rangle = \cos \theta |0\rangle +e^{i\phi}\sin\theta|1\rangle$, and the Bloch sphere of coordinates $\{x,y,z\}$ in this way,
\begin{eqnarray}
\left\{
\begin{array}{ccc}
x&=& 2j =2 \cos\theta\sin\theta\cos\phi\\
y&=& 2h=2\cos\theta\sin\theta\sin\phi\\
z&=& 2k-1=2 \cos ^2\theta-1.
\end{array}
\right.
\end{eqnarray}
It is immediate to verify that the following constraint holds
\begin{eqnarray}
k^2+h^2+j^2=k.
\end{eqnarray}
The algorithm originally designed for $X$--states in \cite{alber}  is flawed  just in not considering the mutual dependence of $h,j,k$, resulting reliable only for a more restricted class of states identified in \cite{nuovodiscord}.

The above mapping enables us to re-parameterize the conditional entropy, \eq{scond}, as a  function of the  azimuthal and polar angles $\theta,\phi$;  we can then write $\tilde {\cal S}(h,j,k)=\tilde {\cal S}(\theta,\phi)$ and  perform the optimization of $\tilde {\cal S}$ over these two independent variables.

\subsection{Optimization}
  Inspired by \cite{alber},  we look for symmetries in the expression of the conditional entropy.   We notice immediately the invariance  under the transformation $\theta\rightarrow \theta \pm \pi$, which, however, can be englobed  by  the following one
 \begin{eqnarray}
 \left\{
  \begin{array}{c}
 k \rightarrow  1-k\\
 h \rightarrow  -h\\
 j \rightarrow  -j,\\
  \end{array}
 \right .
 \end{eqnarray}
  which corresponds to
  \begin{eqnarray}
  \theta \rightarrow \theta \pm \pi/2.
    \end{eqnarray}

\begin{figure}[tb]
\centering \includegraphics[width=8cm ]{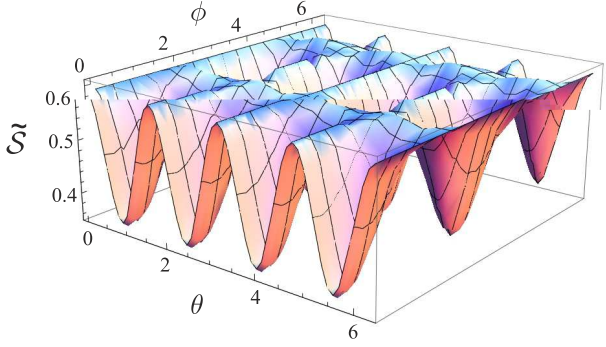}
\caption{(Color online) Example of conditional entropy $\tilde {\cal S}$ for a random two--qubit state [\eq{random}].  The angles $\theta$ and $\phi$ parametrize the measurement: we can appreciate the symmetry properties of $\tilde {\cal S}$ with respect to such variables, expressed by the invariance $\tilde {\cal S}(\theta,\phi)=\tilde {\cal S}(\theta \pm\pi/2, \phi)$. All the quantities plotted are dimensionless.}
\label{discordsample}
\end{figure}

     We can appreciate this with an example.  Let us pick a random state, such as
\begin{equation}\label{random}
{\scriptsize
\left(\!
\begin{array}{cccc}
0.437 & 0.126 + 0.197 i & 0.0271 - 0.0258 i &  -0.274 + 0.0997 i \\
 0.126 -
   0.197 i &
  0.154 &  -0.0115 - 0.0187 i &  -0.0315 +
   0.170 i\\
 0.0271 + 0.0258 i & -0.0115 + 0.0187 i &
  0.0370 &
  0.00219 - 0.0367 i\\
 -0.274 - 0.0997 i &  -0.0315 -
   0.170 i & 0.00219 + 0.0367 i & 0.372
   \end{array}\!
\right)};
 \end{equation}
 we can operate with local unitaries on it, obtaining a new state $\rho$ (albeit with the same entropies and discord) described by the simplified normal form presented in \eq{simpler}; we then perform  a projective measurement on subsystem $B$, obtaining an ensemble whose conditional entropy  is plotted in Fig.~\ref{discordsample}.

One sees that there are no further apparent symmetries for the conditional entropy. Therefore the analysis so far, while not being conclusive, allows us just to refine the problem by safely letting  the optimization of the conditional entropy to be restricted to the interval $\theta \in [0,\pi/2)$. To determine the minimum of $\tilde {\cal S}$, we need to calculate its derivatives with respect to $\theta$ and $\phi$. The dependence on these variables involves  logarithms of non--linear quantities, so we can not expect to solve analytically the problem in any case, whatever ingenious variables we might choose. However, we can seek to write the two constraints in a compact and elegant form. Let us impose
 \begin{eqnarray}
  \left\{
  \begin{array}{ccc}
 p&=&\vec{b}\cdot\vec{X}\\
 r_+&=&|\vec{m}_+ |\\
 r_-&=&|\vec{m}_-|;
  \end{array}
  \right.
 \end{eqnarray}
After a bit of algebra, we obtain
 \begin{eqnarray}
 \tilde {\cal S}&=&-\frac14\bigg\{(1 - p - r_-) \log_2[1 - p - r_-] \\\nonumber
&+& (1 - p + r_-) \log_2[ 1 - p + r_-] \\ \nonumber &+& (1 + p + r_+) \log_2[ 1 + p + r_+] \\\nonumber
&+&(1 + p - r_+) \log_2[ 1 + p - r_+] \\ \nonumber &-& 4 + 2 (-(1 - p) \log_2[ 1 - p] \\\nonumber
&-& (1 + p) \log_2[ 1 + p])\bigg\} .
 \end{eqnarray}
 Now, we set the partial derivatives to zero,
  \begin{eqnarray}
  \left\{
  \begin{array}{c}
\frac{\partial \tilde {\cal S}}{\partial\theta}= \frac{\partial \tilde {\cal S}}{\partial p}\frac{\partial p}{\partial \theta}+\frac{\partial \tilde {\cal S}}{\partial r_+}\frac{\partial r_+}{\partial \theta}+ \frac{\partial \tilde {\cal S}}{\partial r_-}\frac{\partial r_-}{\partial \theta}=0\,;\\\nonumber \\ \nonumber
\frac{\partial \tilde {\cal S}}{\partial\phi}= \frac{\partial \tilde {\cal S}}{\partial p}\frac{\partial p}{\partial \phi}+\frac{\partial \tilde {\cal S}}{\partial r_+}\frac{\partial r_+}{\partial \phi}+ \frac{\partial \tilde {\cal S}}{\partial r_-}\frac{\partial r_-}{\partial \phi}=0\,.
\end{array}
\right .
 \end{eqnarray}
  Defining the following quantities
  \begin{eqnarray}
  \alpha&=&\det \left(
\begin{array}{cc}
\frac{\partial p}{\partial\theta} &\frac{\partial p}{\partial\phi}\\
 \frac{\partial r_+}{\partial\theta}&\frac{\partial r_+}{\partial\phi}
    \end{array},
\right),\nonumber\\
  \beta&=&\det \left(
\begin{array}{cc}
\frac{\partial p}{\partial\theta} &\frac{\partial p}{\partial\phi}\\
 \frac{\partial r_-}{\partial\theta}&\frac{\partial r_-}{\partial\phi}
    \end{array}
\right),\\
\gamma&=&\det \left(
\begin{array}{cc}
\frac{\partial r_+}{\partial\theta} &\frac{\partial r_+}{\partial\phi}\\
 \frac{\partial r_-}{\partial\theta}&\frac{\partial r_-}{\partial\phi}
    \end{array}\nonumber
\right),
\end{eqnarray}
  after some manipulations, we can write the stationarity conditions in the following form
 \begin{eqnarray}
  \left\{
  \begin{array}{c}
 \frac 14 \log_2 [\frac{1+p-r_+}{1+p+r_+}]+\frac 12 \log_2[\frac{(1+p)(1-p-r_-)}{(1-p)(1+p-r_+)}]\frac{\beta}{\alpha+\beta+\gamma}=0\,;\\ \\
\frac 14 \log_2 [\frac{1-p-r_-}{1-p+r_-}]-\frac 12 \log_2[\frac{(1+p)(1-p-r_-)}{(1-p)(1+p-r_+)}]\frac{\alpha}{\alpha+\beta+\gamma}=0\,.
 \end{array}
\right .
 \end{eqnarray}

We see immediately  that this system can be further simplified to
\begin{eqnarray}
  \left\{
  \begin{array}{c}
 \frac{\log_2 [\frac{1+p-r_+}{1+p+r_+}]}{\beta}+ \frac{\log_2 [\frac{1-p-r_-}{1-p+r_-}]}{\alpha}=0\,;\\ \\
 \frac 14 \log_2 [\frac{1-p-r_-}{1-p+r_-}]-\frac 12 \log_2[\frac{(1+p)(1-p-r_-)}{(1-p)(1+p-r_+)}]\frac{\alpha}{\alpha+\beta+\gamma}=0\, .
  \end{array}
\right .
 \end{eqnarray}

We can still express these equations as relations among the eigenvalues of the ensemble $\{p_k,\rho_k\}$. Calling $\lambda_0^{+},\lambda_0^-$ the eigenvalues of $\rho_0$ and $\lambda_1^+,\lambda_1^-$ the eigenvalues of $\rho_1$, we have
\begin{equation}
\begin{split}
\lambda_0^{\pm}&=\frac12\left(1\pm\frac{r_-}{1-p}\right)\,,\\
\lambda_1^{\pm}&=\frac12\left(1\pm\frac{r_+}{1+p}\right)\,,
\end{split}
\end{equation}
 After some straightforward algebra, one can show that the vanishing of the derivatives of  $\tilde {\cal S}$ occurs when the following constraints are satisfied
  \begin{eqnarray}\label{eletrash}
  \left\{
  \begin{array}{c}
 \lambda_0^{-}=\frac{\left(\frac{\lambda_1^{+}}{\lambda_1^{-}}\right)^{\frac{\alpha}{\beta}}}
 {1+\left(\frac{\lambda_1^{+}}{\lambda_1^{-}}\right)^{\frac{\alpha}{\beta}}}\,;\\ \\
 \lambda_1^{-}= \lambda_0^{-}\left(\frac{\lambda_0^{+}}{\lambda_0^{-}}\right)^{\frac{\alpha+\beta+\gamma}{2\alpha}}\,.
 \end{array}
\right .
 \end{eqnarray}

These two transcendental equations can be solved numerically. They represent the most compact formulation to date for the problem of calculating the quantum discord of arbitrary two--qubit states. Let us call $s_i$ the solutions obtained, corresponding to values $\{\theta_i,\phi_i\}$. In order to establish if they represent minima of $\tilde {\cal S}$, we adopt the conventional method and  evaluate the signature of the Hessian matrix $H$  at the points $\{\theta_i,\phi_i\}$, and, in case of   $\text{det} H =0$, we study the sign of the functions $\delta_i=\tilde {\cal S}(\theta,\phi)-\tilde {\cal S}(\theta_i,\phi_i)$. Naming $\{\theta_{mj},\phi_{mj}\}$ the angles such that $H$ is positive definite or $\delta_i  > 0$,  we clearly have that the absolute minimum of the conditional entropy is defined as
\begin{equation*}
\text{min}_{\{P_{Bk}\}}\sum_k p_k {\cal S}(A|B_{\{P_{Bk}\}})= \text{min}_{\{\theta_{mj},\phi_{mj}\}}\tilde {\cal S}(\theta_{mj},\phi_{mj}).
\end{equation*}
The quantum discord for generic two--qubit states of the form \eq{simpler} finally reads
\begin{equation}\label{discordfinal}
{\cal D}(A:B)= {\cal S}(B)- {\cal S}(A,B) + \text{min}_{\{\theta_{mj},\phi_{mj}\}}\tilde {\cal S}(\theta_{mj},\phi_{mj}).
\end{equation}

 \section{Comparison between quantum discord and geometric discord}\label{sec3}

 We can use our results to compare the quantum discord ${\cal D}$ \cite{zurek} with the geometric discord $D_G$ \cite{dakic} for general two--qubit states \cite{comparezap}.
 We have generated up to $10^6$ random general two--qubit states. After transforming each of them into the normal form of \eq{simpler}, we have calculated their quantum discord ${\cal D}$, as numerically obtained from the algorithm of Section \ref{sec2}, and their normalized geometric discord $2 D_G$. The latter admits the following explicit analytic expression for states $\rho$ in normal form, derived from \eq{dgformula},
 \begin{eqnarray}
  D_G (\rho)= \frac 14(||\vec b \vec b^T||_2 + ||\vec c \vec c^T||_2 -\tilde k),
 \end{eqnarray}
 where $\tilde k$ is the largest eigenvalue of the matrix $\vec b \vec b^T + \vec c \vec c^T$.

\begin{figure}[tb]
\centering
 \includegraphics[width=8.5cm ]{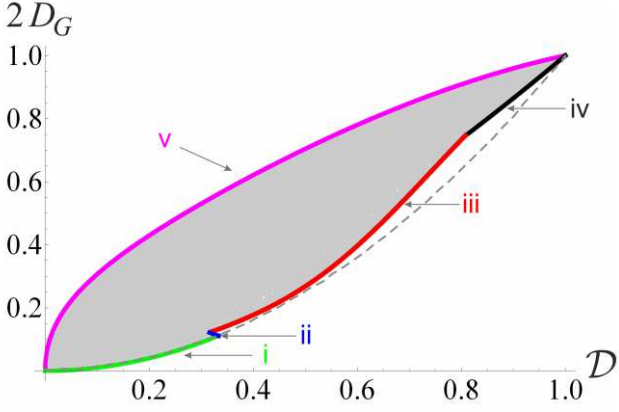}
\caption{(Color online) Comparison between normalized geometric discord $2D_G$ and quantum discord ${\cal D}$ for $10^6$ randomly generated general  two--qubit states. The dashed line is obtained by taking the equality sign in \ineq{hier}. Refer to the main text for details of the other boundary curves. All the quantities plotted are dimensionless.}
\label{zurVSgeo}
\end{figure}

The results are shown in Fig.~\ref{zurVSgeo}. We notice that physical states of two qubits fill up a two-dimensional area in the space of the two non-classicality measures, meaning that the two impose inequivalent orderings on the set of mixed two--qubit quantum states; this is reminiscent of the case of entanglement measures (see e.g. \cite{nemoto}), and a similar feature has been reported concerning the comparison between discord and other non-classicality indicators \cite{gap}. Nevertheless, at fixed quantum discord, the geometric discord admits exact lower and upper bounds (and vice versa). We have identified them numerically.

Firsts of all, we have observed that the following hierarchical relationship holds for arbitrary two--qubit states,
\begin{equation}\label{hier}
2D_G \ge {\cal D}^2\,.
\end{equation}
In other words, the quantum discord for any two--qubit state can never exceed the square root of its (normalized) distance from the set of classical--quantum states. The corresponding boundary curve is plotted as a dashed (gray) line in Fig.~\ref{zurVSgeo}. However, we see that such a bound is tight only in the region $0 \le {\cal D} \le 1/3$, in which it coincides with what we refer to as branch (i) [see below], while it is not attainable for higher degrees of non-classical correlations. The actual, tight lower bound in the whole $\{{\cal D}, 2D_G\}$ plane accommodates states with minimal geometric discord at fixed quantum discord, or equivalently maximal quantum discord at fixed geometric discord: Such extremal states are constituted by the union of four different families, which sit on branches (i)--(iv) in Fig.~\ref{zurVSgeo}.

\begin{enumerate}
  \item[(i)] (Green online) This branch is filled by so-called $\alpha$ states \cite{james}
  \begin{equation}\label{alphastates}
\rho_\alpha=  \left(
\begin{array}{cccc}
 \frac{\alpha }{2} & 0 & 0 & \frac{\alpha }{2} \\
 0 & \frac{1-\alpha }{2} & 0 & 0 \\
 0 & 0 & \frac{1-\alpha }{2} & 0 \\
 \frac{\alpha }{2} & 0 & 0 & \frac{\alpha }{2}
\end{array}
\right)\,,\quad 0 \le \alpha \le 1/3\,,
\end{equation}
for which
${\cal D}(\rho_\alpha) = \alpha$ and $2D_G(\rho_\alpha) = \alpha^2$,
thus saturating \ineq{hier}.
  \item[(ii)] (Blue online)
  This small branch is filled by a subclass of the two-parameter family
  \begin{equation}\label{arstates}\begin{split}
&  \rho_{r}=\left(
\begin{array}{cccc}
 (1-a)/2 & 0 & 0 & r/2 \\
 0 & a & 0 & 0 \\
 0 & 0 & 0 & 0 \\
 r/2 & 0 & 0 & (1-a)/2
\end{array}
\right)\,, \\ &\frac13 \le a \le \frac{5}{14}\,,\quad \sqrt{4a-3a^2-1}\,,\end{split}\end{equation}
with $r \in \left[ \sqrt{4 a-3 a^2-1},\, \frac{1-a}{3}\right]$ given by the solution to $\frac{2 r \tanh ^{-1}\left(\sqrt{a^2+r^2}\right)}{\sqrt{a^2+r^2}}+\ln (-a-r+1)-\ln (-a+r+1)+2 \tanh ^{-1}(r)=0$. The geometric discord of these states is simply $2D_G(\rho_{r})=a^2$, while their quantum discord is calculated in \cite{james}.
  We highlight the presence of the ``pimple'' at the joint between branches (i) and (ii), a recurring feature in the profile of extremal states involving quantum discord \cite{james,galve,gap}.
  \item[(iii)] (Red online) This branch accommodates asymmetric $X$--states of the form
  \begin{equation}\label{asymx}\begin{split}
  \rho_g=\left(
\begin{array}{cccc}
 a & 0 & 0 & \sqrt{a-a^2-a c} \\
 0 & c & 0 & 0 \\
 0 & 0 & 0 & 0 \\
 \sqrt{a-a^2-a c} & 0 & 0 & 1-a-c
\end{array}
\right)\,, \\ a = \frac{1-2 c+2 c^2-g}{2 c}\,, \,\,0\le g \le 1\,,\end{split}
\end{equation}
with $$c \in \left[\frac{1-\sqrt{g}}2,\, \frac12 - \left\{
                                                \begin{array}{ll}
                                                  \frac12\sqrt{2g-1}, & g>\frac12; \\
                                                  0, & \hbox{otherwise;}
                                                \end{array}
                                              \right.\right]\,,
$$
solution to $8 (1-2 c) c^2 \tanh ^{-1}\left(\sqrt{8 (c-1) c-2 g+3}\right)-4 c^2 \sqrt{8 (c-1) c-2 g+3} \tanh ^{-1}(1-2 c)+2 \sqrt{8(c-c^2)-2 g+3} \left(2 c^2+g-1\right) \tanh ^{-1}\left(\frac{3c-2 c^2 +g-1}{c}\right)=0$. For these states, $2D_G(\rho_g)=g$ and ${\cal D}(\rho_g) = \frac{1}{\ln 4}\, \bigg[-\ln (-4 c (a+c-1))-2 \sqrt{4 c (a+c-1)+1} \tanh ^{-1}\left(\sqrt{4 c (a+c-1)+1}\right)-2 \ln (1-a)+4 a \tanh ^{-1}(1-2 a)+2 \ln (2-2 c)-4 c \tanh ^{-1}(1-2 c)\bigg]$.

  \item[(iv)] (Black online) The top-right-most branch accommodates just pure states $\rho_p=\ket{\psi}_{AB}\bra{\psi}$, for which the discord equals the marginal von Neumann entropy,   ${\cal D}(\rho_p)={\cal S}(\rho_A)=-p \log_2 p - (1-p) \log_2 (1-p)$, and the geometric discord equals the marginal linear entropy, $2D_G(\rho_p) = 2(1-\text{Tr} \rho_A^2) = 4 \det \rho_A=4p(1-p)$, where we have denoted the eigenvalues of the reduced density matrix $\rho_A$ by $\{p, 1-p\}$.
\end{enumerate}

\smallskip
\noindent
On the other hand, the upper boundary (v) in Fig.~\ref{zurVSgeo}, despite being single--branched, is more involved and we are unable to provide a tractable parametrization of the states that saturate it. They can be sought among symmetric $X$--states of full rank, but with the two biggest eigenvalues dominating the other two. The extremal curve has been obtained as the result of extensive numerical optimization, in which the parameter space has been finely sliced in discrete intervals of nearly constant discord, and for each interval the datapoint corresponding to the random state with the maximum geometric discord has been selected. Joining all such extremal states we have obtained the smooth (Magenta online) line of Fig.~\ref{zurVSgeo}.

The two measures ${\cal D}$ and $2D_G$ correctly coincide on classical--quantum states \eq{cq}, where both vanish, and on maximally entangled Bell states, where both reach unity.

\section{Conclusions}\label{sec4}
We have presented a reliable and effective algorithm for the evaluation of the quantum discord ${\cal D}$ of general two--qubit states.  We have simplified the optimization involved in calculating the conditional entropy, by removing the redundant degrees of freedom that can be set to zero by means of local unitaries in the first place, and by properly taking into account the symmetries of the problem. The optimization problem for the conditional entropy, and equivalently for the discord, is recast into a compact form that implies an elegant relationship among the eigenvalues of the ensemble obtained after the local measurement process on one qubit. The derived transcendental constraints are amenable to direct numerical solution.

We have then compared quantum discord  with an alternative but affine quantity, the geometric discord $D_G$, identifying the classes of states with extremal values of geometric discord at fixed quantum discord.  For a fixed geometric discord, maximal quantum discord is attained by different families of states depending on the degree of non-classical correlations, encompassing pure as well as mixed, symmetric and nonsymmetric states. In general, the hierarchical bound ${\cal D} \le \sqrt{2 D_G}$ holds for all two--qubit states.

We hope that our results can provide further insight into the fascinating but still not well-understood paradigm of non-classical correlations beyond entanglement in composite quantum systems. The methods presented here can be generalized to higher--dimensional and continuous variable systems.

\acknowledgments{We thank Mauro Paternostro and Steve Campbell for very fruitful exchanges during the early stages of this project.}


\begin{thebibliography}{99}

\bibitem{hororev} R. Horodecki,  P. Horodecki, M. Horodecki, and K. Horodecki, Rev. Mod. Phys. {\bf 81}, 865 (2009).


\bibitem{nielsen}
M.~A. Nielsen and I.~L. Chuang, \emph{Quantum Computation and Quantum
  Information}\ (Cambridge University Press, Cambridge, 2000).


\bibitem{dattaqc} A. Datta, S. T. Flammia, and C. M. Caves,
Phys. Rev. A \textbf{72} 042316 (2005);
A. Datta, and G. Vidal, Phys. Rev. A {\bf 75}, 042310 (2007);
A. Datta, A. Shaji, and C. M. Caves,
Phys. Rev. Lett \textbf{100}, 050502 (2008);  B. P. Lanyon, M. Barbieri, M. P. Almeida, and A. G. White, Phys. Rev. Lett. {\bf 101}, 200501 (2008).

\bibitem{piani}
M. Piani,
P. Horodecki, and R. Horodecki,
Phys. Rev. Lett. \textbf{100}, 090502 (2008).

\bibitem{singapore}
M. Piani, S. Gharibian, G. Adesso, J. Calsamiglia, P. Horodecki, and A. Winter, arXiv:1103.4032 (2011).


\bibitem{zurek} H. Ollivier and W. H. Zurek, Phys. Rev. Lett. {\bf 88}, 017901 (2001).
\bibitem{HV} L. Henderson and V. Vedral, J. Phys. A {\bf 34}, 6899 (2001).

\bibitem{terhal} B. M. Terhal, M. Horodecki, D. W. Leung, and D. P.DiVincenzo,  J. Math. Phys. {\bf 43}, 4286 (2002); D. P. DiVincenzo, M. Horodecki, D. Leung, J. Smolin, and B. M. Terhal,  Phys. Rev. Lett. {\bf 92}, 067902 (2004).

\bibitem{mid} S. Luo, Phys. Rev. A {\bf 77}, 022301 (2008).

\bibitem{deficit} A. K. Rajagopal and R. W. Rendell, Phys. Rev. A {\bf 66}, 022104 (2002).

\bibitem{moelmer} S. Wu, U. V. Poulsen, and K. M{\o}lmer, Phys. Rev. A {\bf 80}, 032319 (2009).

\bibitem{modi} K. Modi, T. Paterek, W. Son, V. Vedral, and M. Williamson, Phys. Rev. Lett. {\bf 104}, 080501 (2010).

\bibitem{dakic} B. Daki\'c, C. Brukner, and V. Vedral, Phys. Rev. Lett. {\bf 105}, 190502 (2010).

\bibitem{gap} D. Girolami, M. Paternostro, and G. Adesso, arXiv:1012.4302 (2010).

\bibitem{rossignoli}R. Rossignoli, N. Canosa, and L. Ciliberti, Phys. Rev. A {\bf 82}, 052342 (2010).

\bibitem{nonclasscorr}
L. Mandel, Phys.
Scr. {\bf T12}, 34 (1986); G. M. D'Ariano,
M. F. Sacchi, and P. Kumar, Phys. Rev. A 59, 826 (1999);
W. Vogel, Phys. Rev. Lett. {\bf 84}, 1849 (2000);
Th. Richter and W. Vogel,
Phys. Rev. Lett. {\bf 89}, 283601 (2002); J. K. Asb\'oth, J. Calsamiglia, and H. Ritsch, Phys. Rev. Lett. {\bf 94},  173602 (2005).


\bibitem{usediscord}  R. Dillenschneider, Phys. Rev. B {\bf 78}, 224413 (2008);
C. A. Rodriguez-Rosario, K. Modi, A. Kuah, A. Shaji, and E. C. G. Sudarshan, J. Phys. A: Math. Theor. {\bf 41}, 205301 (2008);
A. Shabani and D. A. Lidar, Phys. Rev. Lett. {\bf 102}, 100402 (2009);
M. S. Sarandy, Phys. Rev. A {\bf 80}, 022108 (2009);
T. Werlang, S. Souza, F. F. Fanchini, and C. J. Villas-Boas, Phys Rev. A {\bf 80}, 024103 (2009);
B. Bylicka, and D. Chruscinski,  Phys. Rev. A {\bf 81}, 062102 (2010);
L. Mazzola, J. Piilo, and S. Maniscalco, Phys. Rev. Lett. {\bf 104}, 200401 (2010);
L. C. Celeri, A. G. S. Landulfo, R. M. Serra, and G. E. A. Matsas, Phys. Rev. A {\bf 81}, 062130 (2010);
A. Brodutch, and D. R. Terno,  Phys. Rev. A {\bf 81}, 062103 (2010);
M. D. Lang and C. M. Caves, Phys. Rev. Lett. {\bf 105}, 150501 (2010);
T. Werlang, C. Trippe, G. A. P. Ribeiro, and G. Rigolin, Phys. Rev. Lett. {\bf 105}, 095702 (2010);
F. F. Fanchini, L. K. Castelano, and A. O. Caldeira, New J. Phys. {\bf 12}, 073009 (2010);
B. Wang, Z. Xu, Z. Chen, M. Feng, Phys. Rev. A {\bf 81}, 014101 (2010);
F. F. Fanchini, T. Werlang, C. A. Brasil, L. G. E. Arruda, and A. O. Caldeira, Phys. Rev. A. {\bf 81}, 052107 (2010);
D. O. Soares-Pinto, L. C. Celeri, R. Auccaise, F. F. Fanchini, E. R. deAzevedo, J. Maziero, T. J. Bonagamba, and R. M. Serra, Phys. Rev. A {\bf 81}, 062118 (2010);
J.-S. Xu, X.-Y. Xu, C.-F. Li, C.-J. Zhang, X.-B. Zou, and G.-C. Guo, Nat. Commun. {\bf 1}, 7 (2010);
K. Bradler, M. M. Wilde, S. Vinjanampathy, and D. B. Uskov, Phys. Rev. A {\bf 82}, 062310 (2010);
A. Datta, arXiv:1003.5256 (2010); M. F. Cornelio, M. C. de Oliveira, and F. F. Fanchini, arXiv:1007.0228 (2010);
K. Modi, M. Williamson, H. Cable, and V. Vedral, arXiv:1003.1174 (2010);
B. Eastin, arXiv:1006.4402 (2010); F. F. Fanchini, M. F. Cornelio, M. C. de Oliveira, A. O. Caldeira, arXiv:1006.2460 (2010); A. Brodutch and D. R. Terno, Phys. Rev. A {\bf 83}, 010301 (2011).


\bibitem{usemid} A. Datta and S. Gharibian, Phys. Rev. A {\bf 79}, 042325 (2009); A. Datta, Phys. Rev. A {\bf 80}, 052304 (2009); R. Srikanth, S. Banerjee, and C. M. Chandrashekar, Phys. Rev. A {\bf 81}, 062123 (2010); A. Auyuanet and L. Davidovich, Phys. Rev. A {\bf 82}, 032112 (2010).


\bibitem{ferraro}
A. Ferraro, L. Aolita, D. Cavalcanti, F. M. Cucchietti, and A. Acin
 Phys. Rev. A {\bf 81}, 052318 (2010).

\bibitem{discordGauss} G. Adesso and A. Datta, Phys. Rev. Lett. {\bf 105}, 030501 (2010);  P. Giorda and M. G. A. Paris, {\it ibid.}  {\bf 105}, 020503 (2010).

\bibitem{operdiscord} W. H. Zurek, Phys. Rev. A 67, 012320 (2003); D. Cavalcanti, L. Aolita, S. Boixo, K. Modi, M. Piani, and A. Winter,
   Phys. Rev. A {\bf 83}, 032324 (2011); V. Madhok and A. Datta, Phys. Rev. A {\bf 83}, 032323 (2011); A. Streltsov, H. Kampermann, and D. Bruss, Phys. Rev. Lett. {\bf 106}, 160401 (2011).

\bibitem{mista}J. Maziero, L. C. Celeri, and R. Serra, arXiv:1004.2082 (2010).

\bibitem{provapovm}S. Hamieh, R. Kobes, and H. Zaraket, Phys. Rev. A {\bf 70}, 052325 (2004).

\bibitem{luo} S. Luo, Phys. Rev. A {\bf 77}, 042303 (2008).

\bibitem{alber} M. Ali, A. R. P. Rau, and G. Alber, Phys. Rev. A {\bf 81}, 042105 (2010); see also Erratum, Phys. Rev. A {\bf 82}, 069902(E) (2010).

\bibitem{cinesi_sensati}X.-M. Lu,  J. Ma, Z. Xi, and X. Wang,  Phys. Rev. A {\bf 83}, 012327 (2011).

\bibitem{nuovodiscord}Q. Chen, C. Zhang, S. Yu, X.X. Yi, and C. H. Oh, arXiv:1102.0181 (2010).

\bibitem{james} A. Al-Qasimi and D. F. V. James, Phys. Rev. A {\bf 83}, 032101 (2011).

\bibitem{galve} F. Galve, G. L. Giorgi, and R. Zambrini, Phys. Rev. A {\bf 83}, 012102 (2011).

\bibitem{luofu} S. Luo and S. Fu, Phys. Rev A, {\bf 82}, 034302 (2010).

\bibitem{noterespect}  In this respect we remark, as already pointed out in \cite{cinesi_sensati}, that the most used analytical method in literature, introduced in \cite{alber} for the special subclass of $X$--states,  is not completely reliable, since it does not take into account all the constraints concerning the variables that characterize the measurement. This problem is overcome in our formulation.


\bibitem{nemoto}
T.-C. Wei, K. Nemoto, P. M.  Goldbart, P. G.  Kwiat, W. J. Munro, and F. Verstraete, Phys. Rev. A {\bf 67}, 022110 (2003).

\bibitem{verstraete}
F. Verstraete, J. Dehaene, and B. De Moor, Phys. Rev. A {\bf 64}, 010101(R) (2001).

\bibitem{comparezap}
A similar study has been recently attempted in [J. Batle, A. Plastino, A. R. Plastino, and M. Casas, arXiv:1103.0704 (2011)], albeit without identification of the extremal states.


\end{thebibliography}
\end{document}